\documentclass[aps,print,twocolumn,superscriptaddress]{revtex4-1}
\usepackage{graphicx}
\usepackage{dcolumn}
\usepackage{bm}
\usepackage{epsfig}
\usepackage{natbib}
\usepackage{amsmath}
\usepackage{amssymb}
\usepackage{times}
\usepackage{psfrag}
\usepackage[english]{babel}
\usepackage[normal]{subfigure}

\begin{document}

\title{Entanglement Manipulation  in a Quantum Chaotic Optical Fiber by Modifying its Geometry}
\author{Sijo K. Joseph,}
\email[]{sijo.joseph@urjc.es}
\affiliation{Nonlinear Dynamics, Chaos and Complex Systems Group, Departamento de F\'{i}sica, Universidad Rey Juan Carlos, Tulip\'{a}n s/n, 28933 M\'{o}stoles, Madrid, Spain.}

\author{Juan Sabuco}
\affiliation{Nonlinear Dynamics, Chaos and Complex Systems Group, Departamento de F\'{i}sica, Universidad Rey Juan Carlos, Tulip\'{a}n s/n, 28933 M\'{o}stoles, Madrid, Spain.}

\author{Lock Yue Chew}
\affiliation{Division of Physics and Applied Physics, School of Physical and Mathematical Sciences, Nanyang Technological University, 21 Nanyang Link, Singapore 637371}

\author{Miguel A. F. Sanju\'{a}n }
\affiliation{Nonlinear Dynamics, Chaos and Complex Systems Group, Departamento de F\'{i}sica, Universidad Rey Juan Carlos, Tulip\'{a}n s/n, 28933 M\'{o}stoles, Madrid, Spain.}

\date{\today}

\begin{abstract}
The effect of boundary deformation on the non-separable entanglement which appears in the classical electromagnetic field is considered. A quantum chaotic billiard geometry is used to explore the influence of a mechanical modification of the optical fiber cross-sectional geometry on the production of non-separable entanglement within classical fields.  For the experimental realization of our idea, we propose an optical fiber with a cross section that belongs to the family of Robnik chaotic billiards. Our results show that a modification of the fiber geometry from a regular to a chaotic regime can enhance the transverse mode entanglement. Our proposal can be realized  in a very simple experimental set-up which consists of a specially designed optical fiber where non-entangled light enters at the input end and entangled light propagates out at the output end after interacting with a fiber boundary that is known to generate chaotic behavior.
\end{abstract}
 
 \maketitle

\section{Introduction}

Recent research in quantum chaos has uncovered the relationship between chaos and entanglement
both  theoretically and experimentally \cite{Chaudhury09,Ghose_Art04,Wang_Ent_Indi,skj_physlett}. It is well known that classically chaotic trajectories exhibit rich topological structures in the classical phase space \cite{ChaosTopologyBook}. Recently, there has been an increasing interest in relating quantum entanglement and classical geometry. In this article, we explore how quantum-like entanglement can be intimately connected to the geometry of the domain boundary within which the wavefunctions are confined. In this context, a quantum chaotic billiard in the non-relativistic quantum regime is best suited for this study, with the effects of spatial geometry on entanglement and its connection to Hamiltonian chaos to be an as yet unexplored direction.
Classical chaos has been observed in very simple systems like the Robnik billiard \cite{Robnik_Billiard} and the Bunimovich stadium\cite{Bunimovich_Stadium}. It is already known that a quantum system in a classically chaotic regime can generate higher entanglement\cite{Bandyopadhyay02,Bandyopadhyay04,Miller99,Fujisaki03,Chaudhury09,Lombardi11,Arul01,Zhang08,Chung09}. In terms of its implementation, a recent experiment has demonstrated the presence of quantum chaos in light beams \cite{ChaoLight}.  While quantum entanglement has been practically implemented in sophisticated discrete-variable systems such as an NMR based system \cite{Manu1}, there is a recent interest in the application of continuous variable entanglement  for the purpose of quantum computation \cite{LloydPRL}. The approach based on the analogies between quantum physics and optics via the classical electromagnetic waves is particularly relevant in this context as in \cite{SimonPRL}. It was Spreeuw who realized the existence of non-quantum optical entanglement \cite{Spreeuw1,Spreeuw2}. According to Ghose, the EPR like correlation can also appear in classical electrodynamics \cite{ParthaGhose1,ParthaGhose3}. More recently it is shown that the classical electrodynamics contains a Hilbert space structure which makes it similar to quantum mechanics in certain aspects \cite{ParthaGhose2}. In the recent experiments, researchers have explored the non-quantum entanglement in the classical optical field \cite{Xiangdong,Eberly}. It is well known that the Maxwell's equations of the transverse modes of the electric field in an optical fiber can be transformed into an Optical-Schr\"odinger equation, where the $z$-axis of the optical fiber is analogous to time \cite{analogsquantum1,analogsquantum2,analogsquantum3,analogsquantum4,china_analog1}. Hence, the classical electromagnetic wave equation  can be used to mimic the quantum features and facilitate a proper experimental implementation. H. J. St\"ockmann et al. has already explored this type of analogy between electromagnetic waves and quantum mechanics  to explore quantum chaos using  a two-dimensional microwave cavity \cite{Stockmann1,Stockmann2,Stockmann3,stockmannbook}. Here, we use geometries inspired from studies of quantum chaos to explore the geometric dependence of entanglement. We introduce the idea of exploiting an optical fiber with a core cross-section that has the geometry of a chaotic billiard to create a highly entangled light beam.
More specifically,  we analyze in this paper an optical fiber which possesses the cross sectional geometry of the Robnik billiard. By analyzing the propagation of a classical electromagnetic wave in these fibers, we can explore the geometric dependence of the non-separable entanglement in terms of the lowest-order eigenmodes, as well as using  coherent and squeezed coherent states as the initial wavepackets.

\section{Two-dimensional Optical-Schr\"odinger equation}
It is well known that a light beam propagating in an optical fiber can be described by the following equation,
\begin{eqnarray}
i\lambda_{w} \frac{\partial\psi(x,y,z)}{\partial z}&=&-\frac{\lambda_{w}^2}{2n_{0}(z)}{\nabla_{xy}}^2\psi(x,y,z) \nonumber \\
&&+\frac{1}{2n_{0}(z)}{[{n_{0}}^2(z)-{n}^2(x,y,z)]}\psi(x,y,z) \nonumber \\
\label{waveeq}
\end{eqnarray}
which is a two-dimensional Optical-Schr\"odinger equation with an optical-wavefunction in the transverse $x$ and $y$ variables. The longitudinal coordinate $z$ along the fiber axis is similar to time in the standard Schr\"odinger equation  \cite{analogsquantum1,analogsquantum2,analogsquantum3,china_analog1}. This type of analogy between quantum mechanics and wave optics has been also pointed out in many other works \cite{QMlike_Optics,Opt_Entangle,Exper_OptEnt}.
Here $n_{0}(z) $ is the refractive index along the beam axis and $\lambda_{w}$ is the wavelength of the electromagnetic wave. Hence, the optical quantum Hamiltonian can be written as,
\begin{eqnarray}
\hat{H}=\frac{1}{n_{0}(z)}\left( \frac{{\hat{p}_{x}^2}}{2}+\frac{{\hat{p}_{y}}^2}{2}\right)+U(x,y,z)
\end{eqnarray}
and the optical potential function  has the form $U(x,y,z)= \frac{1}{2n_{0}(z)}{[{n_{0}}^2(z)-{n}^2(x,y,z)]}$. This type of two-dimensional Hamiltonian has already been studied in the context of chaotic ray propagation in an optical fiber \cite{ChaoRayHam}.  In  the usual sense, the classical limit corresponds to $\hbar\to 0$, while in the Optical-Schrodinger equation this limit corresponds to $\lambda_{w}\to 0$ which is just the classical ray optics approximation. By properly choosing the refractive index profile, we can define our potential function.

Let $D$ be the domain of the fiber core cross-section in the two-dimensional $xy$ plane. The optical fiber is a dielectric medium with refractive index $n(x,y,z)$. It is known that for a step-index fiber, $n(x,y,z) = n_{0}$ in the core with $(x,y) \in D\cup \partial D $ and $n(x,y,z) = n_{1}$ in the cladding if $(x,y) \in \bar{D}$. We assume that the outer region of the cladding is sufficiently large so that it can be taken to be infinite when considering guided light in the core and near the core-cladding boundary. Under this situation the components of the monochromatic electric and magnetic fields obey the Helmholtz equation.  In order to study the  time evolution of the wavepacket, we define our two-dimensional potential as a function which is  zero inside the domain (fiber core region) and relatively high outside the domain (cladding region).  Hence our problem can be represented by the Helmholtz equation in a 2D domain which has been employed to describe diverse quantum chaotic billiards. In other words, our problem of the  wavepacket propagation in an optical fiber has been reduced to a quantum chaotic billiard problem. The two-dimensional potential is taken as zero inside the domain and a high value of around $10^{12}$ outside the domain. This is enough to make the wavefunction to vanish on the outer boundary. In order to represent the two-dimensional potential on a grid, the Robnik billiard is represented in cartesian coordinates which is given by
\begin{eqnarray}
((u+B)^2+v^2-2B(u+B))^2=A^2((u+B)^2+v^2). \nonumber \\
\end{eqnarray}
Using the cartesian form, the region inside and outside the Robnik billiard can be easily determined. Once the two-dimensional potential is obtained numerically, the wavefunction is propagated via the split operator method. Then, the density matrix is computed to evaluate the von Neumann entropy of entanglement $S_{vn}$.

\begin{figure}[!]
\begin{center}
\includegraphics[width=3in,height=2in]{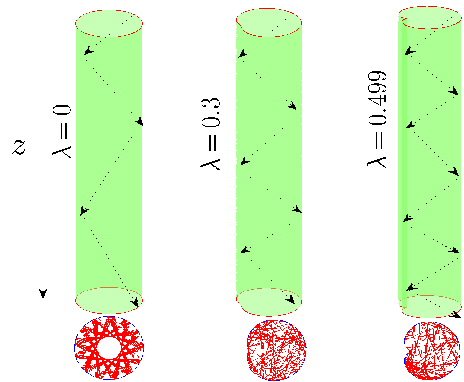}
\caption{This figure shows the geometry at the core of a family of the Robnik optical fiber. When $\lambda=0$, the core of the optical fiber takes the standard circular shape. By increasing the deformation parameter $\lambda$, different quantum chaotic optical fibers are obtained. A schematic of the optical ray propagating in the $z$-direction is also shown as well as its projection on the $xy$ plane which contains patterns reminiscent of a quantum chaotic billiard. Note that the $z$ axis is the analog of time in the standard Schr\"odinger equation.}\label{NeatFiber}
\end{center}
\end{figure}

In the classical limit, we study the dynamics of the classical billiard via ray optics while the analogous quantum mechanics represents the wave optics description in terms of the electromagnetic wave and its propagation in the waveguide. The quadrature components $x$, $y$, $\hat{p}_{x}$ and $\hat{p}_{y}$ are the position and momentum coordinate  in the plane perpendicular to the fiber axis. This is clarified in Fig.~\ref{NeatFiber} where we have shown the wave propagation in the optical fiber and its projection on the $(x,y)$ plane. It is observed that the projection on the $(x,y)$ plane is a quantum chaotic Robnik billiard. Note that in our case $x$, $y$, ${p}_{x}$ and ${p}_{y}$ are exactly the coordinates and momentum variables of the quantum chaotic Robnik billiard. In the ray optics limit, the path of light rays illustrate the chaotic properties while in the wave optics regime, the interference of light waves gives rise to quantum-like effects.

\section{Mechanical modification of the quantum entanglement}
The goal of this article is to elucidate the effect of the domain boundary geometry on non-separable entanglement.  Our results would illustrate the possibility of manipulating  optical entanglement via modifying the mechanical design of the core geometry. We shall entirely focus on the Robnik billiard due to its relative simplicity in terms of numerical computations. Our idea can be intuitively grasped from the following fact that the eigenfunctions of a rectangular billiard is formed by a tensor product of $x$ and $y$ modes, and hence it is not entangled. By slightly deforming the rectangular geometry to a Bunimovich stadium, the eigenfunction would be changed entirely. The new eigenfunctions of the Bunimovich stadium then becomes a linear combination of the eigenfunctions of the rectangular billiard with proper boundary conditions and it is no longer a product state. The same is true for the Robnik billiard which results from deforming a circle. The eigenfunctions of the circular billiard \cite{circ_eigen} can be expressed as a tensor product state in terms of the $r$ and $\theta$  variables as follows:
\begin{eqnarray}
\phi_{m,n}(r,\theta)=J_{m}({j_{mn}r)} e^{im\theta}.
\end{eqnarray}
In the radial direction, the eigenmodes are described by the Bessel functions $J_{m}({j_{mn}r)}$, while the polar coordinates give the wavefunction $\Theta=e^{+im\theta}$
and  $j_{mn}$ describes the $n$th zero of the Bessel functions of order $m$.
Hence, there is no entanglement between the polar and the radial part of the wavefunction for the circular geometry. By deforming the geometry, quantum entanglement starts to appear. For example, the eigenfunction of the Robnik billiard can be expanded in terms of the eigenfunctions of the circular billiard as follows:
\begin{eqnarray}
\psi(\mathbf{r,\theta})=\sum_{m=0}^{N}\sqrt{C_{m}}{{J_{m}}{(\beta r)}e^{im\theta}}.\label{schimdt}
\end{eqnarray}
The value of $\beta$ can be determined by minimizing the norm of the wavefunction on the boundary, which then gives the complete information about the wavefunction $\psi(\mathbf{r,\theta})$.  This technique is often used in the numerical implementation of the method of particular solutions \cite{Timo1,Timo2}. Equation~\ref{schimdt} shows that the wavefunction $\psi(\mathbf{r,\theta})$ is in the form of the Schmidt decomposition, where the Schmidt basis are the eigenfunctions of the circular billiard and the entanglement  can be determined easily from the coefficients $C_{m}$. As the deformation increases, we need more terms to approximate the eigenfunction. Hence, the entanglement entropy would increase. When the beam paths configuration becomes highly chaotic, a larger number of Schmidt modes are required to approximate the eigenfunction which leads to a higher entanglement.  This results from the fact that the field modes start to interact through the special geometry of the deformed circular boundary. This simple observation has far reaching consequences since it implies the possibility of manipulating entanglement via mechanical methods which is technically easy to achieve. In addition, quantum chaos would serve to indicate the critical geometry that gives a maximum entanglement in the system.

In order to  study the continuous variable entanglement of the transverse mode $\psi(\mathbf{r,\theta})$ in quantum chaotic optical fibers, we first obtain the reduced density matrix of the first subsystem $\rho_{1}$ by integrating over the field variable $\mathbf{\theta}$ in terms of the bipartite wavefunction $\psi(\mathbf{r,\theta})$ as follows:

\begin{equation}
\rho_{1}(\mathbf{r,{r}^{'}})=\int {\psi(\mathbf{r,\theta}){\psi}^{*}(\mathbf{{r}^{'},\theta}) \sqrt{rr^{'}}d\theta\;}.
\end{equation}

To quantify the continuous variable entanglement, we compute the von Neumann entropy of the entanglement by using the numerical methods proposed by Parker et al. \cite{ParkBose_Svn} and Bogdanov et al. \cite{Bogdanov}. The von Neumann entanglement entropy of the reduced density matrix is given by,
\begin{equation}
S_{vn}(t)=-\sum\eta_{i}\log \eta_{i} \label{svn_eq}
\end{equation}
where $\eta_{i}$ are  the eigenvalues of the Hermitian kernel $\rho_{1}(\mathbf{r,{r}^{'}})$ as shown in Eq.~\ref{fredholm_eq} below.
The eigenvalues are computed from the Fredholm type II integral equation of $\rho_{1}(\mathbf{r,{r}^{'}})$, which is given by
\begin{equation}
\int {\rho_{1}(\mathbf{r,{r}^{'}})\phi_{i}(\mathbf{{r}^{'}}) d{r}^{'}}=\eta_{i}\;\phi_{i}(\mathbf{r}), \label{fredholm_eq}
\end{equation}
with $\phi_{i}(\mathbf{r})$ being the corresponding Schmidt eigenfunction.
\begin{figure}[!]
\begin{center}
\subfigure[]{\includegraphics[width=3in,height=2in]{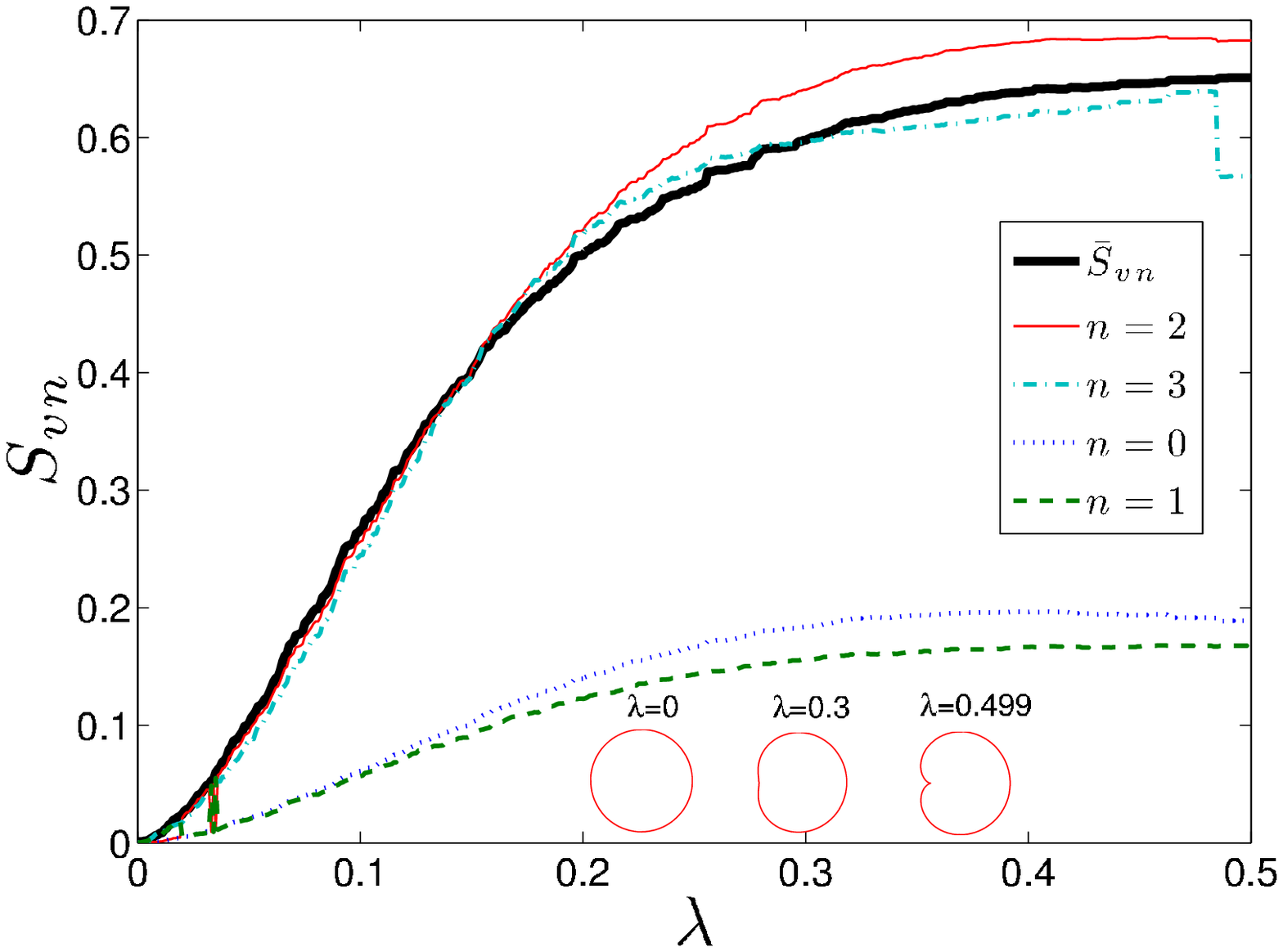}}
\subfigure[]{\includegraphics[width=3in,height=2in]{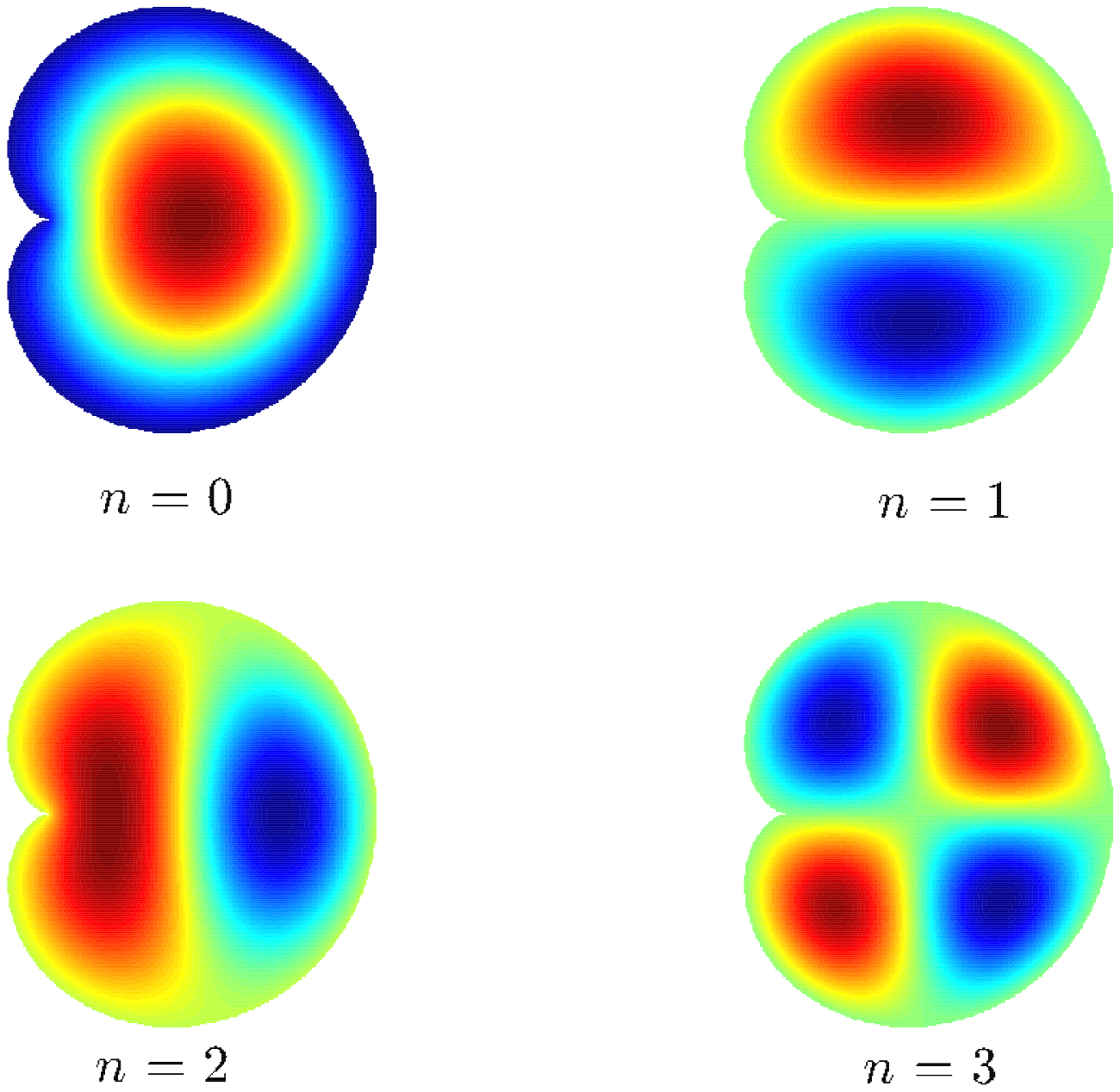}}
\caption{The von Neumann entropy of entanglement $S_{vn}$ of the four lowest eigenfunctions vs the deformation parameter $\lambda$ of the Robnik fiber is shown in (a), while (b) shows the four lowest state eigenfunctions with $\lambda=0.499$. It can be seen that the von Neumann entropy of  entanglement $S_{vn}$ in the eigenmodes saturates as the geometries of these fibers approach the completely chaotic regime. The solid black curve in (a) shows the average von Neumann entropy of entanglement of the ten lowest eigenmodes $\bar{S}_{vn}$ for different boundary geometries of the Robnik fiber.}\label{EntEigen}
\end{center}
\end{figure}
\begin{figure}[]
\begin{center}
\subfigure[] {\includegraphics[width=3in,height=2in]{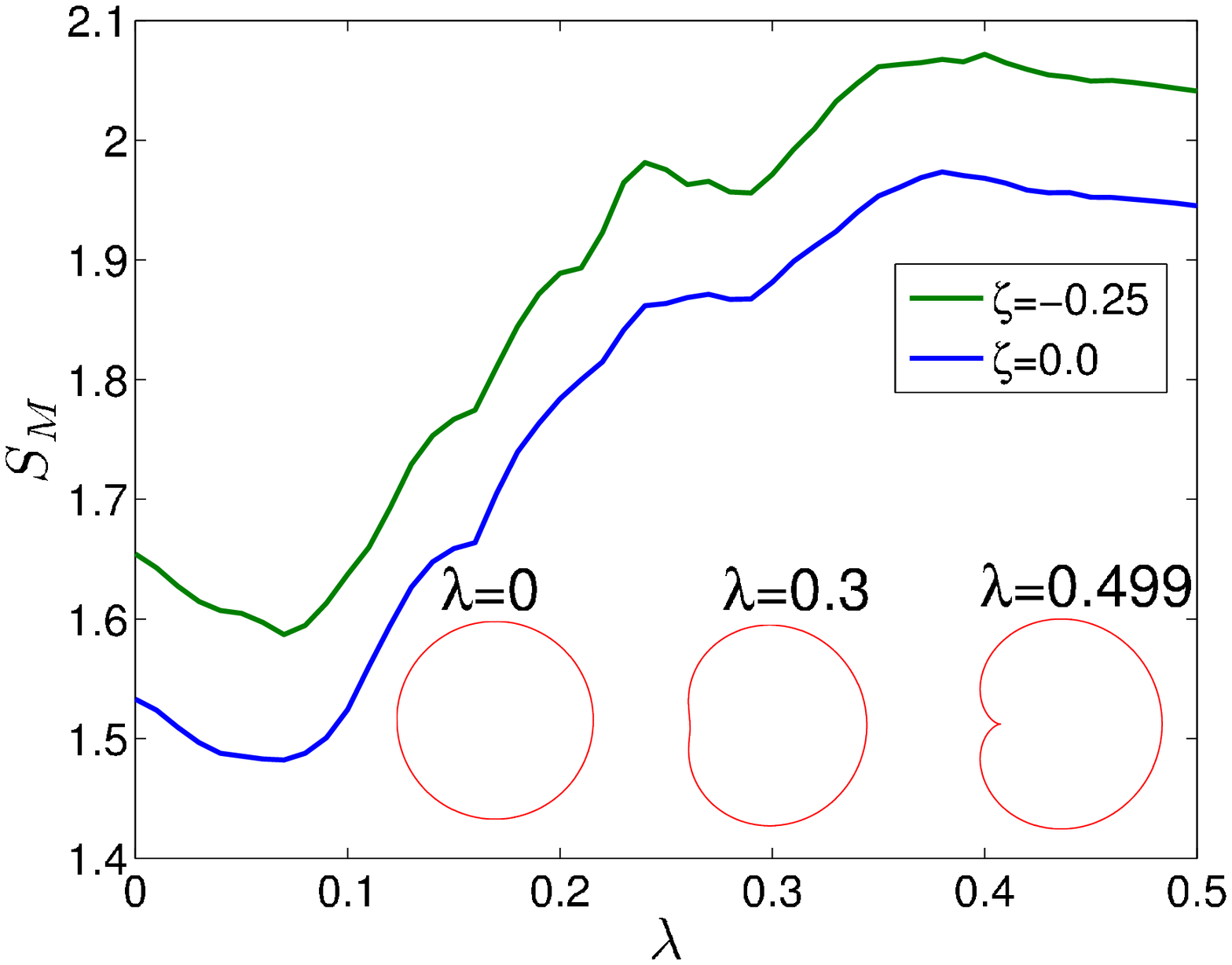}}
\subfigure[] {\includegraphics[width=3in,height=2in]{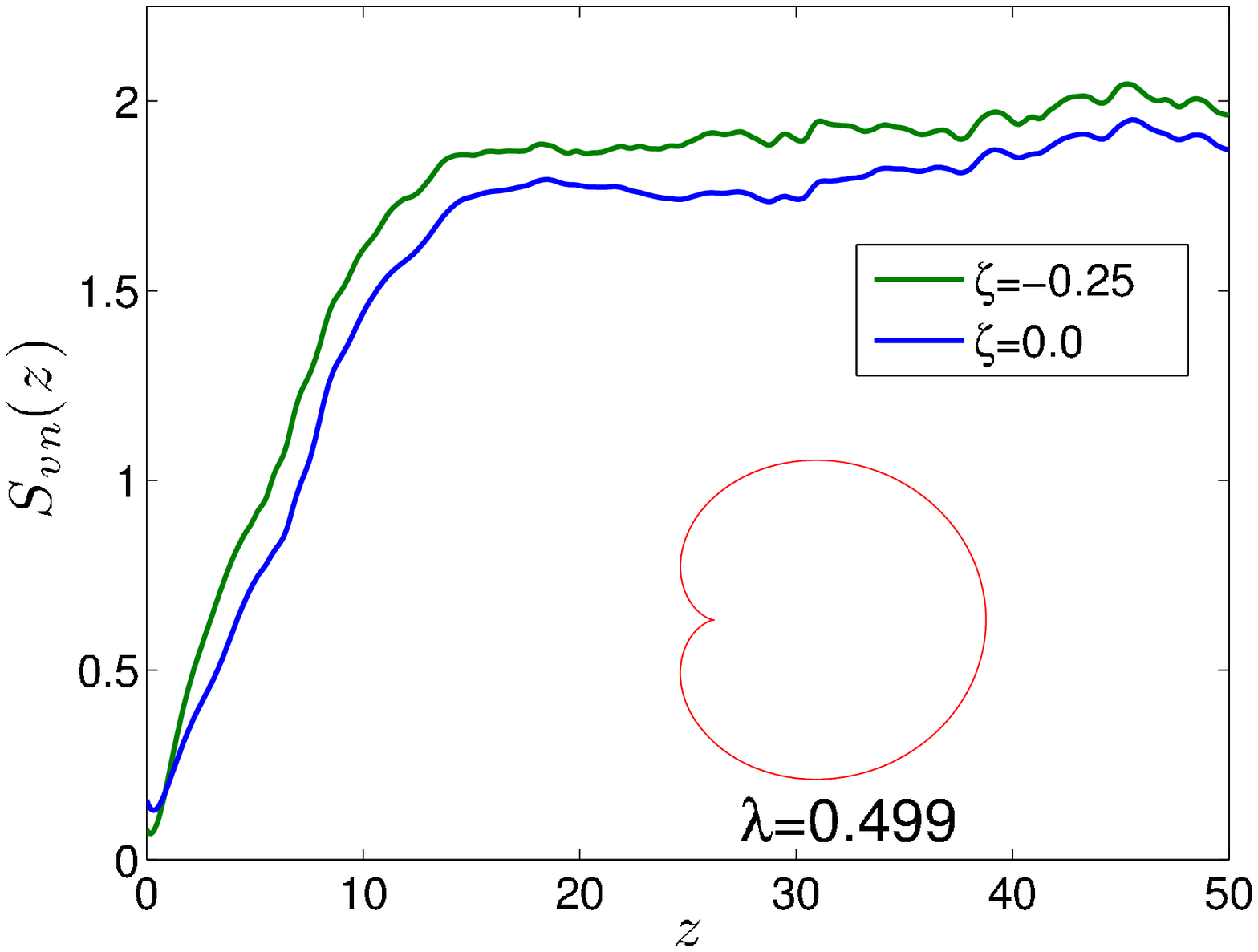}}
\caption{ A plot of the entanglement of the squeezed coherent state versus the deformation parameter $\lambda$ of the Robnik fiber is shown in (a). We observe that the entanglement is higher when the system is in the chaotic regime and the initial squeezing has enhanced the entanglement in the system. In (b), we have plotted the entanglement dynamics as the light propagates along the $z$-axis. It is important to note that the light enters the fiber without any entanglement. Entanglement increases as it propagates, and as it traverses a distance of $50$ units the entanglement saturates. Note that the blue and green curves represent the coherent state and the squeezed coherent state respectively. We have chosen the initial state at the point $(x,p_{x},y,p_{y})=(0.25,0.1,0.0,0.1)$. }
\label{entmax_dyn}
\end{center}
\end{figure}
\begin{figure*}[!htb]
\begin{center}
\includegraphics[scale=0.75,bb=-70 0  600 500]{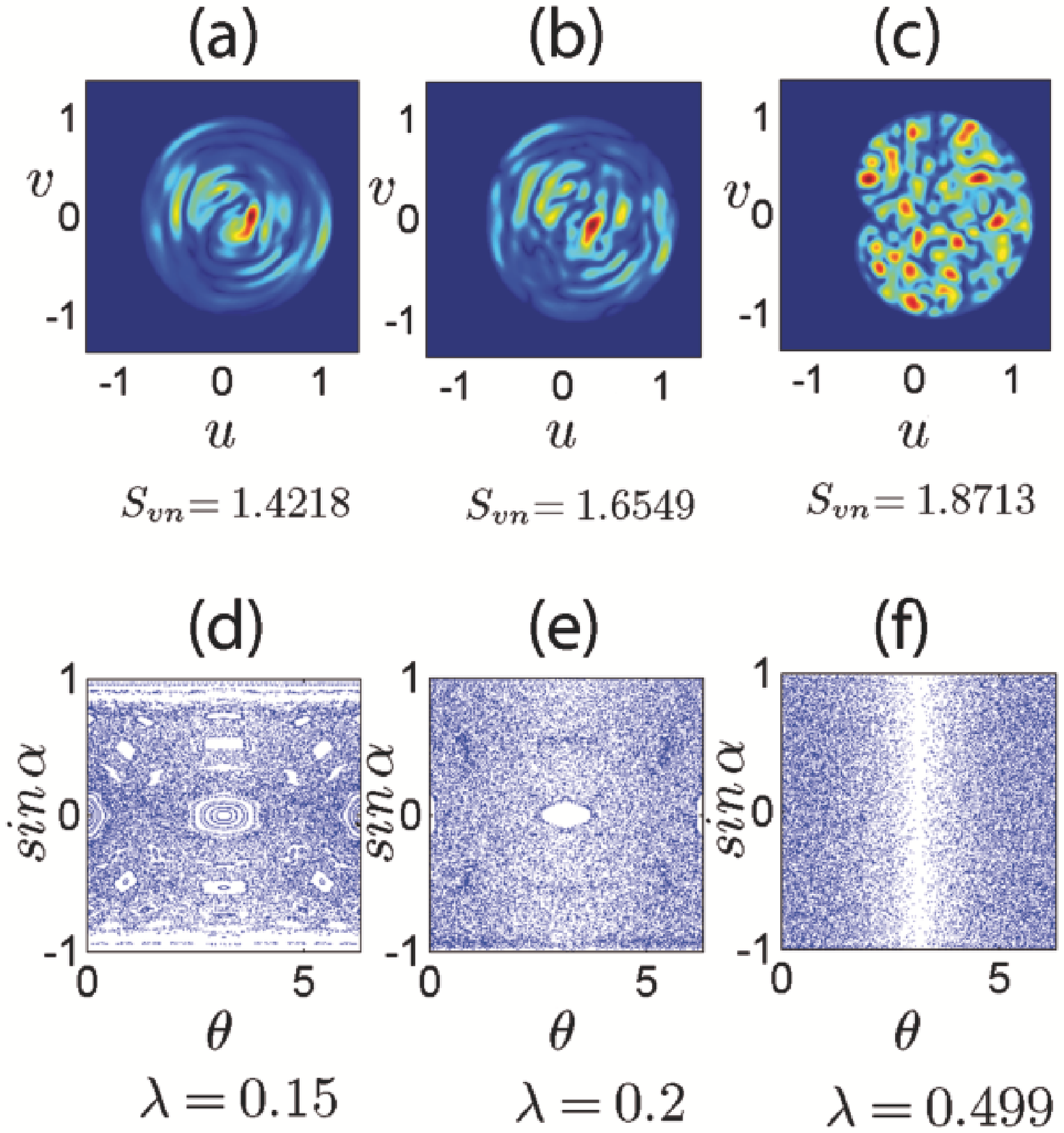}
\caption{The probability densities $|\psi(\mathbf{r,\theta})| ^2$ of the entangled coherent states coming out of the Robnik fiber with length $z=50$ units are shown in (a), (b) and (c) with $\lambda=0.15$, $\lambda=0.2$ and $\lambda=0.499$ respectively. The probability densities $|\psi(\mathbf{r,\theta})| ^2$ are shown in the cross-sectional $u-v$ plane of the fiber. The blue color shows the regions where the probability is zero and the red color shows the regions of higher probability. The entangled state is obtained from a tensor product coherent state centered at $(x,p_{x},y,p_{y})=(0.25,0.1,0.0,0.1)$ with $\lambda_{w}=0.01$. The entangled state occurs after the light beam has propagated a distance of $z = 50$ units. Note that we have also indicated the von Neumann entanglement entropy $S_{vn}$ of the polar coordinates below these figures.  In (d), (e) and (f) the corresponding classical phase space is shown for $\lambda=0.15$,$\lambda=0.2$ and $\lambda=0.499$ respectively. From the figure it can be observed that the time evolved coherent state has a higher entanglement when the corresponding geometry possesses a larger degree of chaotic behavior.}\label{Grid_wave}
\end{center}
\end{figure*}
\section{Robnik Fiber and Entanglement}

The cross section of a Robnik fiber has the shape of the Robnik billiard which is shown in Fig.~\ref{NeatFiber}.
As light passes though the Robnik fiber the transverse modes of the fiber get entangled. We are interested in the optimum geometry which gives a high entanglement in the system. The cross sectional geometry of the Robnik fiber is a conformal transformation $w=A Z (\theta)+B Z(\theta)^2$ of the unit circle $Z(\theta)$ on the complex plane. According to Robnik \cite{Robnik_Billiard}, the cross-sectional geometry can be written in the parametric form as follows,
\begin{eqnarray}
u=A\cos \theta+B\cos 2\theta \\
v=A\sin \theta+B\sin 2\theta .
\end{eqnarray}
The area of the billiard is given by $Area=\pi(A^2+2B^2)$ and by means of a reparametrisation, $A$ and $B$ can be written as follows: $A=\cos p$ and $B=\frac{1}{\sqrt{2}}\sin p$, where $p$ is given by $p=\tan^{-1}{(\lambda\sqrt{2})}$. This allows us to define a family of closed analytical boundaries of constant area, $Area=\pi$, which depends on the values of the deformation parameter $\lambda$. Robnik had shown that the parameter $\lambda$ can be chosen within the interval $0\leq p \leq {p}_{sing}=\tan^{-1}{(\frac{1}{\sqrt{2}})}$ . For smaller values of $\lambda$, the billiard exhibits a regular dynamics since it is closer to the circular geometry. An increase in the parameter $\lambda$ shifts the system from regular to chaotic. We analyze the dependence of the quantum-like entanglement as we vary the deformation parameter $\lambda$ from the regular  to the chaotic regime.
\subsection{Entangled Robnik eigenmodes}
We have varied the fiber cross section by adjusting the deformation parameter $\lambda$. As the boundary geometry changes, we compute the lowest order eigenfunctions, so that the entanglement in the radial $r$ and the angular $\theta$ modes can be evaluated. Based on our earlier discussion, the entanglement in the circular geometry between the $r$ and $\theta$ variable is zero since the eigenfunction is a product state. As we change $\lambda$, the quantum-like entanglement in the system starts to increase.
In Fig.~\ref{EntEigen}(a), we have plotted the entanglement of the eigenmodes against different values of  the deformation parameter $\lambda$.
In the same figure, we have also displayed the corresponding boundary geometry.  We observe from Fig.~\ref{EntEigen}(a) that the entanglement dynamics increases as $\lambda$ increases. As the boundary geometry leads to chaotic behavior, the entanglement in eigenmodes increases even further and as the system approaches the highly chaotic regime the entanglement entropy saturates.  Thus, the degree of chaos in the system is directly linked to the level of entanglement in the light pulse albeit there is a maximum attainable level in lieu of the eventual saturation. We have plotted the eigenmodes corresponding to the saturation level at $\lambda=0.499$ in Fig.~\ref{EntEigen}(b). From Fig.~\ref{EntEigen}(a), we observe that the lowest energy mode $n=0$ has higher entanglement compared to the excited mode $n=1$. This can be understood by examining Fig.~\ref{EntEigen}(b), where the ground state eigenmode is seen to be more severely affected by the deformation of the boundary, while the first excited mode $n=1$ contains a nodal line through its center and has a form that is closer to the eigenmodes of the circular cross-section. Hence, the deformation in the eigenfunction is smaller for the $n=1$ mode. The number of basis terms to describe this eigenfunction is smaller, which implies that its entanglement is smaller. The same argument applies to the excited modes $n=2$ and $n=3$, respectively. These results indicate that the eigenfunctions that are affected more by the boundary deformation would have a higher entanglement. It thus provides insights into how quantum chaotic geometry can be effectively used  to produce highly entangled states.
\subsection{Effect of squeezing and entanglement in the Robnik fiber}

While it is typical to propagate a Gaussian beam in an optical fiber, one may wonder the effect of initially squeezing such a beam before launching it into the fiber. This question has led us to investigate the difference in entanglement dynamics with respect to the initial coherent and the squeezed coherent state as the light wave propagates in the horizontal $z$ direction. For our investigation, we use the Hollenhorst and Caves definition of the squeezing operator \cite{Hollenhorst79,JagdishRai88,Caves81} and we employ the  coordinate representation of the squeezed coherent state \cite{Moller96} for the numerical computations.
The squeezed coherent state is defined as
\begin{eqnarray}
|\alpha_{k},\zeta_{k}\rangle & =  &\hat{D}({\alpha}_{k}) \hat{S}({\zeta}_{k}) |0 \rangle \, ,
\end{eqnarray}
where the displacement operator $\hat{D}$ and the squeezing operator $\hat{S}$ is given by
\begin{eqnarray}
\hat{D}(\alpha_{k})& = & \exp{({{\alpha}_{k}} \hat{a_{k}}^{\dagger}-{\alpha}_{k}^* \hat{a_{k}})} \,,\\
\hat{S}(\zeta_{k})& = & \exp{(\frac{1}{2}{\zeta}_{k} \hat{a_{k}}^{\dagger^{2}}-\frac{1}{2}{{\zeta}_{k}}^* \hat{a_{k}}^{2})} \, ,
\end{eqnarray}
and  $\alpha_{k}=|\alpha_{k}|e^{ \mathrm{i} \phi_{k}}$, $\zeta_{k}=|r_{k}|\;e^{ \mathrm{i} \theta_{k}}$ are complex numbers and $\alpha_{k}$ are related to the phase space variables $(q_{k},p_{k})$ in the following manner
\begin{equation}
\alpha_{k}=\frac{1}{\sqrt{2\hbar}}(q_{k}+\mathrm{i}p_{k}),
\end{equation}
with $k=1,2,$  respectively.
According to M{\o}ller  et al. \cite{Moller96}, the squeezed coherent state in the position basis can be written as
\begin{eqnarray}
& \psi(\mathbf{x},\alpha_{k},\zeta_{k}) =   {\left(\frac{1}{\pi\hbar}\right)}^{1/4}{(\cosh{{r}_{k}}+{\mathrm{e}^{\mathrm{i} \theta}}\sinh{{r}_{k}})}^{-1/2} \nonumber \\  & \! \!
\exp{\left\{-\frac{1}{2\hbar}\left(\frac{\cosh{{r}_{k}}-{\mathrm{e}^{\mathrm{i} \theta}}\sinh{{r}_{k}}}{\cosh{{r}_{k}}+{\mathrm{e}^{\mathrm{i} \theta}}\sinh{{r}_{k}}}\right) {(\mathbf{x}-q_{1})}^{2}+\frac{\mathbf{i}}{\hbar}{p_{1}(\mathbf{x}-{q_{1}/2)}} \right\}} \nonumber\\. 
\end{eqnarray}
The tensor product state of this wavefunction is used to study the quantum-like entanglement dynamics for different squeezing parameter values.

In Fig.~\ref{entmax_dyn}(a) the entanglement maximum $S_{M}$ is plotted against different fiber geometries. The entanglement maximum $S_{M}$  is taken from the von Neumann entropy of entanglement $S_{vn}(z)$ which is computed for the propagation of the initial states along the $z$-axis with a fixed geometry.  The initial  tensor product  coherent state and the squeezed coherent state is propagated for a different boundary geometry and we take these initial states to be centered at $(x,p_{x},y,p_{y})=(0.25, 0.1,0.0,0.1)$. We have computed the quantum-like entanglement present in the radial and polar variables of the optical wavefunction. In Fig.~\ref{entmax_dyn}(b) the entanglement dynamics of the coherent state and squeezed coherent state is plotted against the $z$-axis. When the optical wave with a tensor product coherent state enters the fiber (at $z=0$), its entanglement is zero. During its propagation along the $z$-axis the transverse modes get entangled and after traveling a distance of $50$ units along the $z$-axis, the entanglement becomes saturated.  In other words, we observe that a non-entangled light enters at one end of the fiber and comes out entangled at the other end. In the case of a highly quantum chaotic fiber with $\lambda=0.5$, this entanglement can yield a higher value. Note that we have chosen the initial condition $(x,p_{x},y,p_{y})=(0.25, 0.1,0.0,0.1)$ because it lies entirely inside the domain for the chosen parameter range of $\lambda$.  The wavelength (analog of $\hbar$) given in Eq.~\ref{waveeq} is selected as $\lambda_{w}=0.01$. The value of $\lambda_{w}$ is set small enough in order for the wavepacket to be contained entirely inside the defined two-dimensional cross-sectional domain. We have also used the equally position squeezed $\zeta_{1}=\zeta_{2}=\zeta$ wavepacket at a relatively small value of $\zeta=-0.25$ due to the numerical considerations. From Fig.~\ref{entmax_dyn} we can clearly see that as the core boundary deformation $\lambda$ increases, the production of entanglement is found to increase for both the case of initial coherent state and initial squeezed coherent state. This is again observed in Figs.~\ref{Grid_wave}(a), \ref{Grid_wave}(b) and \ref{Grid_wave}(c) according to the corresponding von Neumann entropy of entanglement. It is well known that when the deformation parameter approaches $\lambda=0.5$,  the dynamical behavior of the classical system is highly chaotic (see Figs.~\ref{Grid_wave}(d), \ref{Grid_wave}(e) and \ref{Grid_wave}(f)) with the higher entanglement maxima observed in the corresponding quantum system. Thus, our results clearly indicate the effect of the geometrical deformation on entanglement dynamics of initial squeezed coherent states in a quantum chaotic optical fiber. It is to be noted that a judicious choice of a quantum chaotic core cross-section would lead to the generation of highly entangled transverse modes. These results are also in accordance with our previous studies \cite{skj_epjd,skj_physlett}. 

\section{Conclusions}
We have explored the dependence of the quantum entanglement with the boundary deformation. We have seen that as the boundary become deformed into the quantum chaotic regime, the entanglement in the eigenmode increases. This demonstrates a simple approach to manipulate entanglement via  mechanical means. We have also analyzed the propagation of a coherent state and a squeezed coherent state in a quantum chaotic Robnik optical fiber. We have found that an initial squeezing can indeed enhance entanglement in quantum chaotic fibers. More importantly, our results have specifically proven the efficiency of manipulating entanglement in an optical fiber by modifying its cross-sectional geometry.

\section{Acknowledgements}
This work was supported by the Spanish Ministry of Science and Innovation under project number FIS2013-40653-P.
\bigskip 
%

\end{document}